\documentstyle[aps,epsfig,floats]{revtex}

\tighten
\renewcommand{\ol}{\overline}
\newcommand{\Pslash}{\kern 0.2 em P\kern -0.56em \raisebox{0.3ex}{/}}
\newcommand{\pslash}{\kern 0.2 em p\kern -0.4em /}
\newcommand{\kslash}{\kern 0.2 em k\kern -0.45em /}
\newcommand{\Sslash}{\kern 0.2 em S\kern -0.56em \raisebox{0.3ex}{/}}
\newcommand{\Mslash}{\kern 0.2 em M\kern -0.70em \raisebox{0.3ex}{/}}
\newcommand{\g}{\gamma}
\newcommand{\sig}{\sigma}
\newcommand{\eps}{\epsilon}
\newcommand{\dg}{\dagger}

\newcommand{\lcvec}[3]{\left[\;#1\;,\;#2\;,\;#3\;\right]}
\newcommand{\sT}{{\scriptscriptstyle T}}
\newcommand{\nn}{\nonumber}
\newcommand{\newangle}{{<\kern -0.3 em{\scriptscriptstyle )}}}
\newcommand{\sumint}{\kern 0.2 em {\textstyle\sum} \kern -1.1 em \int_X}

\begin{document}
\draft
\title{
\begin{flushright}
{\small hep-ph/9907475}
\end{flushright}
Two-hadron interference fragmentation functions \\
Part I: general framework
}

\author{A.~Bianconi$^{1}$, S.~Boffi$^{2}$, R. Jakob$^{2}$, M.~Radici$^{2}$}
\address{
1. Dipartimento di Chimica e Fisica per i Materiali e per l'Ingegneria,\\ 
Universit\`{a} di Brescia, I-25133 Brescia, Italy,\\
2. Dipartimento di Fisica Nucleare e Teorica, Universit\`{a} di Pavia, 
and\\
Istituto Nazionale di Fisica Nucleare, Sezione di Pavia, 
I-27100 Pavia, Italy}

\date{July 23, 1999}

\maketitle

\begin{abstract}
We investigate the properties of interference fragmentation functions 
measurable from the distribution of two hadrons produced in the same jet 
in the current fragmentation region of a hard process. We discuss
the azimuthal angular dependences in the leading order cross 
section of two-hadron inclusive lepton-nucleon scattering as an example 
how these interference fragmentation functions can be addressed separately.
\end{abstract}


\section{Introduction}\label{sec:intro}

Although since the event of Quantum Chromodynamics we do have a 
renormalisable quantum field theory at hand to describe the interaction of 
quarks and gluons, we still have to face the fact that there is no 
rigorous analytical explanation for the confinement of those partons in 
hadrons. For the investigation of the properties of hadron structure we 
mainly rely on the information extracted from experimental data on hard 
scattering processes in form of distribution and fragmentation functions, 
which can be compared to the predictions of different 
models. A complete calculation of these objects from first principles, as 
for instance in lattice gauge theory, is not yet available.

There are three fundamental quark distribution functions (DF) contributing 
to a hard process at leading order in an expansion in powers of the hard 
scale $Q$: the momentum distribution $f_1$, the longitudinal spin 
distribution $g_1$, and the transversity distribution $h_1$. Whereas 
$f_1$ and $g_1$ are experimentally rather well measured, presently the 
transversity distribution $h_1$ is still completely unknown. The reason 
is that it is a chiral-odd object and needs to be combined with a 
chiral-odd partner to form a (chiral-even) cross section. As such, it is not 
measurable, for instance, in totally inclusive deep inelastic 
scattering (DIS)~\cite{jafzeuth}. Together the three fundamental DF 
characterize the state of quarks in the nucleon with regard to the 
longitudinal momentum and to its spin to leading power in $Q$. The 
inclusion of 
effects related to the transverse momentum of quarks inside the nucleon, 
and/or to subleading orders in $Q$, results in a larger number of 
DF~\cite{piet96}. In particular, so called naive time-reversal odd 
(for sake of brevity ``T-odd'') DF arise, in the sense that the 
constraints due to time-reversal invariance cannot be applied because of, 
e.g., soft initial-state interactions~\cite{abm}, or chiral symmetry 
breaking~\cite{adm}, or so-called gluonic poles attributed to
asymptotic (large distance) gluon fields~\cite{vari,danielpiet}.
These DF can describe also a polarization of quarks inside unpolarized 
hadrons. 

Information on hadronic structure, complementary to the one given by the 
DF, is contained in quark fragmentation functions (FF) describing 
the process of 
hadronization. To leading order, those functions give the probabilities to 
find hadrons in a quark. Experimentally known for some species of hadrons 
is only $D_1$, the leading spin-independent FF, which is the direct 
analogue of $f_1$. The basic reason for such a poor knowledge is related 
to the difficulty of measuring more exclusive channels in hard processes 
(such as, e.g., semi-inclusive DIS) and/or collecting data sensitive to 
specific degrees of freedom of the resulting hadrons (transverse momenta,
polarization, etc..). However, a new generation of experiments (including 
both ongoing measurements like HERMES and future projects like 
COMPASS or experiments at RHIC or ELFE) will have a better ability 
in identifying final states 
and will allow for the determination of more subtle effects. In fact, when
partially releasing the summation over final states several FF become 
addressable which are often related to genuine effects due to Final State 
Interactions (FSI) between the produced hadron and the remnants of the 
fragmenting quark~\cite{piet96}. In this context, ``T-odd'' FF naturally 
arise because the existence of FSI prevents constraints from time-reversal 
invariance to be applied to the fragmentation 
process~\cite{vari1,danielpiet}. The 
usefulness of such an investigation can be demonstrated by considering the 
so-called Collins effect~\cite{coll}, where a specific asymmetry 
measurement in the leptoproduction of an unpolarized hadron from a 
transversely polarized target gives access to $h_1$ through the 
chiral-odd ``T-odd'' 
fragmentation function $H_1^{\perp}$, which describes the probability for 
a transversely polarized quark to fragment into an unpolarized hadron. 

The presence of FSI allows that in the fragmentation process there are at
least two competing channels interfering through a nonvanishing phase. 
However, as it will be clear in Sec.~\ref{sec:two}, this is not enough to 
generate ``T-odd'' FF. A genuine difference in the Lorentz structure of 
the vertices describing the fragmentation processes is needed. This poses a 
serious difficulty in modelling the quark fragmentation into one observed 
hadron because it requires the ability of modelling the FSI between the 
hadron itself and the rest of the jet, unless one accepts to give up the 
concept of factorization. Moreover, it was even argued that in this 
situation summing over all the possible final states could average out any 
effect~\cite{jafjitang}.

Therefore, in this article we will discuss a specific situation in the
hadronization of a current quark, namely the one where two hadrons are
observed within the same jet and their momenta are determined. By
interference of different production channels FF emerge which are 
``T-odd'', and can be both chiral even or chiral odd.

For the case of the two hadrons being a pair of pions the resulting FF 
have been proposed as a tool to investigate the transverse spin dependence of 
fragmentation. Collins and Ladinsky~\cite{collad} considered the 
interference of a scalar resonance with the channel of independent 
successive two pion production. Jaffe, Jin and Tang~\cite{jafjitang} 
proposed the interference of $s$- and $p$-wave production channels, where 
the relevant phase shifts are essentially known. In the forthcoming 
paper~\cite{noi}, we will adopt an extended version of the spectator model 
used in Ref.~\cite{spectator} and will estimate the FF in the case of the 
pair being a proton and a pion produced either through non-resonant 
channels or through the Roper ($1440$ MeV) resonance.

This paper is organized as follows. In 
Sec.~\ref{sec:two} the general conditions for generating ``T-odd'' FF in
semi-inclusive DIS are considered which lead naturally to select the
two-hadron semi-inclusive DIS as the simplest case for modelling a 
complete scenario for the fragmentation process. In Sec.~\ref{sec:three}, 
FF are defined for the two-hadron semi-inclusive DIS as projections of a 
proper quark-quark correlation and some relevant kinematics is discussed. 
In Sec.~\ref{sec:four} we give a general method to determine all 
independent leading order two-hadron FF and discuss their symmetry 
properties. As an example, we demonstrate 
in Sec.~\ref{sec:Xsections} that asymmetry measurements in two-hadron 
inclusive lepton-nucleon scattering allow for the isolation of 
the ``T-odd'' FF making use of the angular dependences in the leading 
order cross section. A brief summary and conclusions are given in 
Sec.~\ref{sec:summary}.

\begin{figure}[h]
\begin{center}
\psfig{file=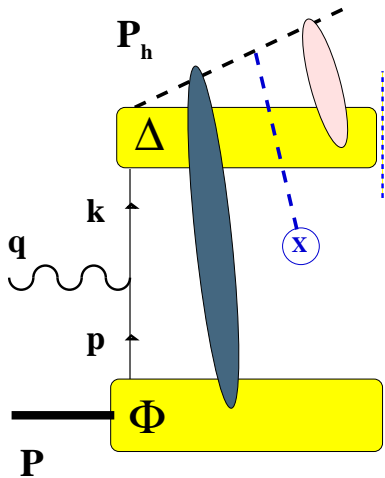, width=6cm}
\end{center}
\caption{Amplitude for a one-hadron semi-inclusive DIS with possible 
residual interactions in the final state of the detected leading 
hadron.}
\label{fig1}
\end{figure}

\section{Why a fragmentation into two hadrons?}
\label{sec:two}

Let us consider the situation of a one-hadron semi-inclusive DIS, where a 
quark with momentum $p$ and carrying a fraction $x$ of the target 
momentum $P$ absorbes a hard photon with momentum $q$ and then hadronizes 
into a jet which contains a leading hadron with momentum $P_h$ eventually 
detected (Fig.~\ref{fig1}). The soft parts that link the initial hard 
quark $p$ to the target and the final hard quark $k$ to the detected 
hadron $P_h$ indicate the distribution probability of the quark itself 
inside the target and the transition probability for its hadronization,
respectively. They are described by the functions $\Phi (p;P,S)$ and 
$\Delta (k;P_h,S_h)$, that may depend also on the initial and final spin 
vectors $S,S_h$ and contain a sum over all possible residual hadronic 
states (symbolized by the vertical dashed lines in Fig.~\ref{fig1}). 

The soft leading hadron can undergo FSI with the sorroundings. In 
Fig.~\ref{fig1} three symbols represent the three possible classes of 
models that could describe them. The dark blob indicates mechanisms (at 
leading or subleading order) that break the factorization hypothesis. The 
light blob represents interactions with the residual fragments. It is 
clear that this second class requires non-trivial microscopic 
modifications of the hadron wave function, in other words it requires the 
ability of modelling the residual interaction between the outgoing hadron 
and the rest of the jet in a way that cannot be effectively reabsorbed in 
the vertex connecting the hard and the soft part. Moreover, it was also 
argued~\cite{jafjitang} that the required sum over all possible states of 
the fragments could average these FSI effects out.

The third symbol, the dashed line originating from the space-time point 
$X$, represents the most naive class of models, where FSI are simply 
described by an averaged external potential. Despite its simplicistic 
approach, this point of view poses serious mathematical difficulties, 
because the introduction of a potential in principle breaks the 
translational and rotational invariance of the problem. One could 
introduce further assumptions about the symmetry properties of the 
potential to keep these features, but at the price of loosing any 
contribution to the ``T-odd'' 
structure of the amplitude, as it will be clear in the following.

\begin{figure}[h]
\begin{center}
\psfig{file=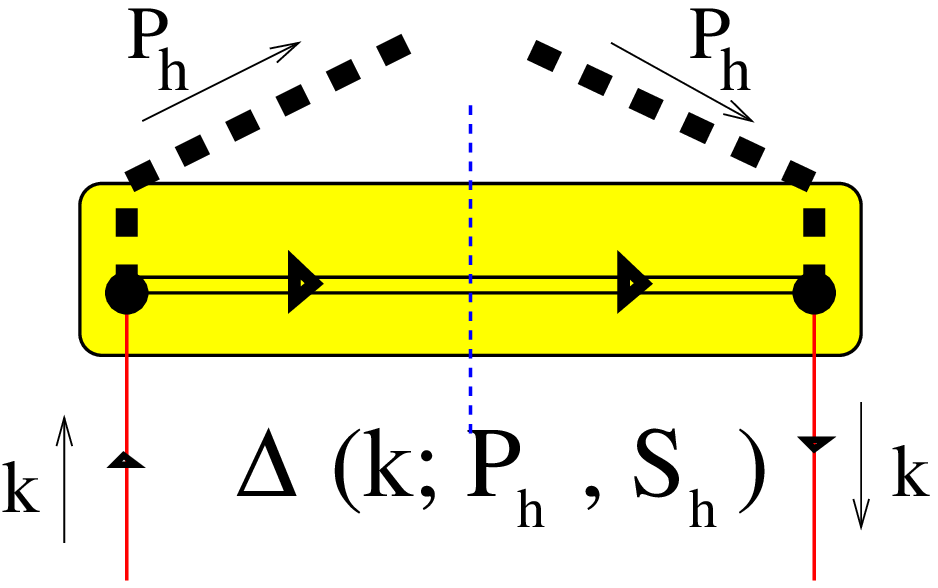, width=7cm}
\end{center}
\caption{Quark-quark correlation function $\Delta$ for the 
fragmentation of a quark into a hadron. A simple model assumption
for the hadronization process is also indicated.}
\label{fig2}
\end{figure}

As a pedagogical example, let us consider the oversimplified and 
unrealistic situation where the detected hadron with mass $M_h$ and energy 
$E_h$ is not polarized and does not interact with the rest of the jet; 
therefore, it is described by the free wave function  
\begin{equation}
\psi_{P_h}(x) = \sqrt{\frac{M_h}{E_h}} \; 
e^{i P_h \cdot x} \; u(P_h) \;,
\label{eq:freespinor}
\end{equation}
where $u(P_h)$ is the free Dirac spinor. In momentum space 
Eq.~(\ref{eq:freespinor}) reads
\begin{equation}
\psi_{P_h}(p) = \frac{1}{(2 \pi)^4} \;  \int d^{\,4\!}p \;  
e^{-i p \cdot x} \; \psi_{P_h}(x) = 
\sqrt{\frac{M_h}{E_h}} \; \delta\left(P_h - p\right) \; u(P_h) .
\label{eq:ftspinor}
\end{equation}
The jet itself is replaced by a spectator system which, again for sake of
simplicity, is assumed to be a structureless on-shell scalar diquark with 
mass $M_D$ and momentum $k-P_h$ in order to preserve momentum conservation 
at the vertex. All this amounts to describe the remnants of the 
fragmentation process with a simple propagator 
line $\delta ((k - P_h)^2 - M_D^2)$ for the point-like on-shell scalar 
diquark $k-P_h$. Then, ignoring 
the inessential $\delta$ functions, the function $\Delta (k;P_h,S_h=0)$ 
of Fig.~\ref{fig2} becomes
\begin{eqnarray}
\Delta (k;P_h, S_h=0) &\sim & \frac{-i}{\kslash-m} \; 
u(P_h)\overline{u}(P_h) \; \frac{i}{\kslash-m}  
\nonumber \\
& = & \frac{\kslash+m}{k^2-m^2} \; (\Pslash_h+M_h) \; 
 \frac{\kslash+m}{k^2-m^2}  \;,
\label{eq:1hsimple}
\end{eqnarray} 
where in the second line the usual projector for two free fermion spinors 
has been used. Eq.~(\ref{eq:1hsimple}) can be cast in the following linear 
combination of all the independent Dirac structures of the process allowed 
by parity invariance~\cite{ralsop,piet96},  
\begin{equation}
\Delta (k;P_h,S_h=0) = A_1 M_h + A_2 \Pslash_h + A_3 \kslash + 
\displaystyle{\frac{A_4}{M_h}} \sigma_{\mu \nu} P_h^{\mu} k^{\nu} ,
\label{eq:1hinv}
\end{equation}
where the amplitudes $A_i$ are given by 
\begin{eqnarray}
A_1 &= & \frac{1}{(k^2 - m^2)^2} 
         \left( k^2 + m^2 + 2 k \cdot P_h \frac{m}{M_h} \right) \nonumber \\
A_2 &= & {}- \frac{1}{k^2 - m^2} \nonumber \\
A_3 &= & \frac{2\left(P_h \cdot k + M_h\, m\right)}{(k^2 - m^2)^2}\nonumber \\
A_4 &= & 0  \;.
\label{eq:1hampls}
\end{eqnarray}
The function $\Delta (k;P_h,S_h=0)$ must meet the applicable constraints 
dictated by hermiticity of fields and invariance under time-reversal 
operations. Hermiticity implies that all $A_i$ amplitudes be real. Since 
the outgoing hadron is described by a free Dirac spinor, time-reversal 
invariance also implies $A_4^* = -A_4$, from which the usual convention in 
naming this amplitude {\it time-odd} (or ``T-odd'') originates. Combining 
the two constraints gives $A_4 = 0$ in agreement with the previous result 
deduced just by simple algebra arguments.

This result holds true if in Eq.~(\ref{eq:ftspinor}) complex momenta are
considered as a simple way to incorporate FSI, as discussed in 
Ref.~\cite{eikonal}. All this amounts to describe the hadron 
wave function as a plane wave damped uniformly in space through the 
imaginary part of $P_h$. The related symmetric potential, or alternatively 
the underlying Dirac structure of free spinor assumed for the hadron wave 
function, do not generate ``T-odd'' structures in the scattering 
amplitude. 

Let us now allow FSI to proceed through a competing channel having a 
different spinor structure with respect to the free channel. This example 
could be cast in the class represented by the light blob of 
Fig.~\ref{fig1}. As a simple test case, we assume for the final hadron 
spinor the following replacement, 
\begin{equation}
u(P_h) \leftrightarrow u(P_h) + e^{i \phi} \kslash u(P_h) ,
\label{eq:dw}
\end{equation}
where $\phi$ is the relative phase between the two channels. Inserting 
this back into Eq.~(\ref{eq:1hsimple}) modifies 
the $\Delta (k;P_h,S_h=0)$ function according to 
\begin{equation}
\Delta (k;P_h,S_h=0) = \left[ A_1 (k^2 +1) + B_1 \cos \phi \right] M_h + 
A_2 (1 - k^2) \Pslash_h + \left[ A_3 + B_3 + B_3' \cos \phi \right] 
\kslash + \frac{B_4 \sin \phi}{M_h} \sigma_{\mu \nu} 
P_h^{\mu} k^{\nu} \;,
\label{eq:1hinvdw}
\end{equation}
where the new amplitudes $B_i$ are given by 
\begin{eqnarray}
B_1 &= & \frac{1}{(k^2 - m^2)^2} \left[ 4\,m\,k^2 + 
\frac{2}{M_h} k \cdot P_h (k^2 + m^2) \right] \nonumber \\
B_3 &= & \frac{1}{(k^2 - m^2)^2} \left( 2\,M_h\,m\,k^2 + 
2\,m^2\,k \cdot P_h \right) \nonumber \\
B_3' &= & \frac{1}{(k^2 - m^2)^2} \left[ 2 M_h (k^2 + m^2) + 
4\,m\,k \cdot P_h \right] \nonumber \\
B_4  &= &\displaystyle{\frac{2 M_h}{k^2 - m^2}}  .
\label{eq:1hamplsdw}
\end{eqnarray}
The coefficient of the tensor structure $\sigma_{\mu \nu}$ is now not 
vanishing provided that the interference between the two channels, namely 
the phase $\phi$, is not vanishing. In this case, a ``T-odd'' contribution 
arises and is maximal for $\phi = \pi /2$. 

These simple arguments show that, in order to model ``T-odd'' FF in 
one-hadron semi-inclusive processes without giving up factorization, one 
needs to relate the modifications of the hadron wave function to a 
realistic microscopic description of the fragmenting jet. Oversimplified 
assumptions, as in the case of symmetric, mathematically handable, 
external potentials, can lead to misleading results. Such a hard task 
suggests that a more convenient way to model occurrence and properties of 
``T-odd'' FF is to look at residual interactions between two hadrons in 
the same jet, considering the remnant of the jet as a spectator 
and summing over all 
its possible configurations. Therefore, in the following the formalism for 
two-hadron semi-inclusive production and FF will be addressed.

\section{Quark-quark correlation function for two-hadron production}
\label{sec:three}

\begin{figure}[h]
\begin{center}
\psfig{file=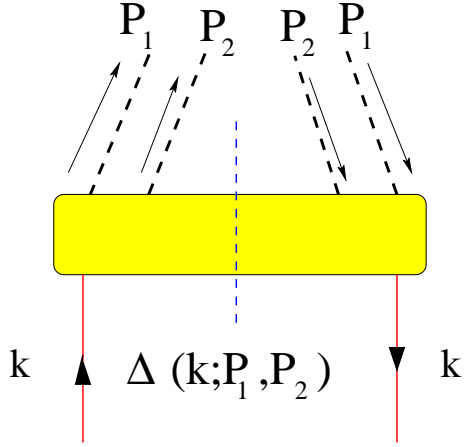, width=6cm}
\end{center}
\caption{Quark-quark correlation function for the fragmentation of a quark 
into a pair of hadrons.}
\label{fig3}
\end{figure}

In the field-theoretical description of hard processes the soft parts
connecting quark and gluon lines to hadrons are defined as certain matrix
elements of non-local operators involving the quark and gluon fields
themselves~\cite{soper,colsop,jaffe}. In 
analogy with semi-inclusive hard 
processes involving one detected hadron in the final state~\cite{piet96}, 
the simplest matrix element for the hadronisation into two hadrons is the 
quark-quark correlation function describing the decay of a quark with 
momentum $k$ into two hadrons $P_1, P_2$ (see Fig.~\ref{fig3}), namely  
\begin{equation}
\Delta_{ij}(k;P_1,P_2)= \sumint\;
\int \frac{d^{\,4\!}\zeta}{(2\pi)^4} \; 
e^{ik\cdot\zeta}\;
\langle 0 \vert\,\psi_i(\zeta)\,a_2^\dg(P_2)\,a_1^\dg(P_1)\,\vert X\rangle \; 
\langle X \vert\,a_1(P_1)\,a_2(P_2)\,\ol{\psi}_j(0)\,\vert 0\rangle \;,
\label{eq:defDelta}
\end{equation}
where the sum runs over all the possible intermediate states involving the 
two final hadrons $P_1,P_2$. For the Fourier transform only the two 
space-time points $0$ and $\zeta$ matter, i.e.\ the positions of quark 
creation and annihilation, respectively. Their relative distance $\zeta$ is 
the conjugate variable to the quark momentum $k$. 

We choose for convenience the frame where the total pair 
momentum $P_h=P_1+P_2$ has no transverse component. The constraint to 
reproduce on-shell hadrons with fixed mass $(P_1^2=M_1^2, P_2^2=M_2^2)$ 
reduces to seven the number of independent degrees of freedom. As shown in 
Appendix A (where also the light-cone components of a 4-vector are 
defined), they can conveniently be reexpressed in terms of the light-cone 
component of the hadron pair momentum, $P_h^-$, of the light-cone fraction 
of the quark momentum carried by the hadron pair, $z_h=P_h^-/k^-=z_1+z_2$, 
of the fraction of hadron pair momentum carried by each individual hadron, 
$\xi=z_1/z_h=1-z_2/z_h$, and of the four independent invariants that can 
be formed by means of the momenta $k,P_1,P_2$ at fixed masses $M_1,M_2$, 
i.e.
\begin{equation}
\tau_h=k^2 \;,\quad
\sig_h = 2\,k\cdot(P_1+P_2)\equiv 2\,k\cdot P_h \;,\quad
\sig_d = 2\,k\cdot(P_1-P_2)\equiv 4\,k\cdot R \;,\quad
M_h^2 = \left(P_1+P_2\right)^2 \equiv P_h^2 \;,
\label{eq:invariants}
\end{equation} 
where we define the vector $R=(P_1-P_2)/2$ for later use.

By generalizing the Collins-Soper light-cone 
formalism~\cite{colsop} for fragmentation into multiple 
hadrons~\cite{collad,jafjitang}, the cross section for two-hadron 
semi-inclusive emission can be expressed in terms of specific Dirac 
projections of $\Delta(z_h,\xi,P_h^-,\tau_h,\sig_h,M_h^2,\sig_d)$ after 
integrating over the (hard-scale suppressed) light-cone component $k^+$ 
and, consequently, taking $\zeta$ as light-like~\cite{piet96}, i.e.
\begin{equation} 
\Delta^{[\Gamma]} = \frac{1}{4z_h}\left.\int dk^+\;\mbox{Tr}[\Delta\Gamma]
\right|_{\zeta^-=0} = \frac{1}{4z_h} \int dk^+\; \int dk^-\; 
\delta\left(k^--\frac{P_h^-}{z_h}\right)\; \mbox{Tr}[\Delta\Gamma] \; .
\label{eq:projDelta}
\end{equation}
The function $\Delta^{[\Gamma]}$ now depends on five variables, apart from 
the Lorentz structure of the Dirac matrix $\Gamma$. In order to make this
more explicit and to reexpress the set of variables in a more convenient 
way, let us rewrite the integrations in Eq.~(\ref{eq:projDelta}) in a 
covariant way using 
\begin{equation}
2\,P_h^-=\frac{d\sig_h}{dk^+},
\qquad
2\,k^+=\frac{d\tau_h}{dk^-} ,
\end{equation} 
and the relation
\begin{equation} 
\frac{1}{2k^+}\delta\left(k^--\frac{P_h^-}{z_h}\right)=
\delta\left(2k^+k^--\frac{2k^+P_h^-}{z_h}\right)=
\delta\left(\tau_h+{\vec k}_\sT^{\,2}-\frac{\sig_h}{z_h}
+\frac{M_h^2}{z_h^2}\right) 
\end{equation} 
which leads to the result 
\begin{equation} 
\Delta^{[\Gamma]}(z_h,\xi,{\vec k}_\sT^{\,2},M_h^2,\sig_d)=
\int d\sig_h \, d\tau_h \;\delta\left(\tau_h+{\vec k}_\sT^{\,2}-
\frac{\sig_h}{z_h}+\frac{M_h^2}{z_h^2}\right)\;\frac{\mbox{Tr}[
\Delta(z_h,\xi,P_h^-,\tau_h,\sig_h,M_h^2,\sig_d) \; \Gamma]}{8z_hP_h^-} 
\;,
\label{eq:projDelta2}
\end{equation}
where the dependence on the transverse quark momentum ${\vec k}_\sT^{\,2}$ 
through $\sig_h$ is made explicit by means of Eqs.~(\ref{eq:sigh2}) and 
(\ref{eq:pt2}). 

\begin{figure}[h]
\begin{center}
\psfig{file=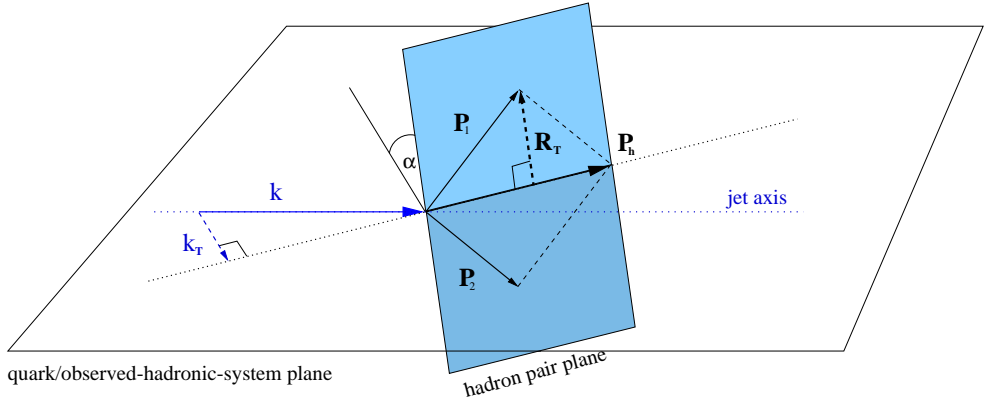, width=13cm}
\end{center}
\caption{Kinematics for a fragmenting quark jet containing a pair of 
leading hadrons.}
\label{fig4}
\end{figure}

Using Eq.~(\ref{explinv2}) makes it possible to reexpress 
$\Delta^{[\Gamma]}$ as a function of $z_h$, $\xi$, ${\vec k}_\sT^{\,2}$ and 
${\vec R}_\sT^{\,2}$,  ${\vec k}_\sT \cdot {\vec R}_\sT$, 
where ${\vec R}_\sT$ is 
(half of) the transverse momentum between the two hadrons in the 
considered frame. In this manner $\Delta^{[\Gamma]}$ depends on how much 
of the fragmenting quark momentum is carried by the hadron pair $(z_h)$, 
on the way this momentum is shared inside the pair $(\xi)$, and on the
``geometry'' of the pair, namely on the relative momentum of the two 
hadrons $({\vec R}_\sT^{\,2})$ and on the relative orientation between the 
pair plane and the quark jet axis (${\vec k}_\sT^{\,2}$, 
${\vec k}_\sT \cdot {\vec R}_\sT$, see also Fig.~\ref{fig4}).

\section{Analysis of interference fragmentation functions}
\label{sec:four}

If the polarizations of the two final hadrons are not observed, the 
quark-quark correlation $\Delta(k;P_1,P_2)$ of Eq.~(\ref{eq:defDelta}) can 
be generally expanded, according to hermiticity and parity invariance, as 
a linear combination of the independent Dirac structures of the process 
\begin{eqnarray}
\Delta(k;P_1,P_2)  &=& B_1\,\left(M_1+M_2\right) + B_2\,\Pslash_1  
+ B_3\,\Pslash_2 + B_4\,\kslash \nn\\[2mm]
&& {}+ \frac{B_{5}}{M_1}\,\sig^{\mu \nu} P_{1\mu} k_\nu
{}+ \frac{B_{6}}{M_2}\,\sig^{\mu \nu} P_{2\mu} k_\nu
{}+ \frac{B_{7}}{M_1+M_2}\,\sig^{\mu \nu} P_{1\mu} P_{2\nu}\nn\\[2mm]
&&
{}+ \frac{B_8}{M_1 M_2}\,\g_5\,\eps^{\mu\nu\rho\sig}
     \g_\mu P_{1\nu} P_{2\rho} k_\sig \;.
\label{eq:ansatz}
\end{eqnarray} 
Symmetry constraints are obtained in the form
\begin{mathletters}
\label{symmetries} 
\begin{eqnarray}
\g_0\,\Delta^\dg(k;P_1,P_2)\,\g_0 &=& \Delta(k;P_1,P_2)
\qquad\qquad\mbox{from hermiticity,} \label{eq:hermt} \\[2mm]
\g_0\,\Delta(\tilde k;\tilde P_1,\tilde P_2)\,\g_0 &=& \Delta(k;P_1,P_2)
\qquad\qquad\mbox{from parity invariance,} \label{eq:parity} \\[2mm]
\left(\g_5C\,\Delta(\tilde k;\tilde P_1,\tilde P_2)\,C^\dg\g_5\right)^\ast 
&=& \Delta(k;P_1,P_2) \quad\qquad\mbox{from time-reversal invariance} \; , 
\label{eq:timerev}
\end{eqnarray} 
\end{mathletters}
where $\tilde a=(a^0,-\vec a)$ and $C=i\,\g^2\,\g^0$. From the hermiticity 
of the fields it follows that 
\begin{equation} 
B_i^* = B_i \qquad\mbox{for}\quad i=1,..,12
\end{equation} 
and, if constraints from time-reversal invariance can be applied, that 
\begin{equation}
B_i^* = B_i \qquad\mbox{for}\quad i=1,..,4
\qquad\qquad
B_i^* = -B_i \qquad\mbox{for}\quad i=5,..,8 \; ,
\label{eq:todd}
\end{equation} 
which means in that case $B_5=B_6=B_7=B_8=0$, i.e.\ terms involving
$B_5,..,B_8$ are naive ``T-odd''.

Inserting the ansatz~(\ref{eq:ansatz}) in Eq.~(\ref{eq:projDelta2}) and
reparametrizing the momenta $k,P_1,P_2$ with the indicated new set of 
variables, we get the following Dirac projections
\begin{eqnarray} 
\Delta^{[\g^-]}(z_h,\xi,{\vec k}_\sT^{\,2},{\vec R}_\sT^{\,2},
   {\vec k}_\sT \cdot {\vec R}_\sT) &\equiv& 
D_1(z_h,\xi,{\vec k}_\sT^{\,2},{\vec R}_\sT^{\,2},
   {\vec k}_\sT \cdot {\vec R}_\sT) \nn\\
&=& \frac{1}{2z_h}\int [d\sig_h d\tau_h] \  \left[ B_2 \,\xi + B_3 \,
     (1-\xi) + B_4 \,\frac{1}{z_h} \right]  \label{eq:projg-} \\[2mm]
\Delta^{[\g^- \g_5]}(z_h,\xi,{\vec k}_\sT^{\,2},{\vec R}_\sT^{\,2},
   {\vec k}_\sT \cdot {\vec R}_\sT) &\equiv& 
\frac{\eps_\sT^{ij} \,R_{\sT i}\,k_{\sT j}}{M_1\,M_2}\;
G_1^\perp (z_h,\xi,{\vec k}_\sT^{\,2},{\vec R}_\sT^{\,2},
   {\vec k}_\sT \cdot {\vec R}_\sT) \nn\\
&=& \frac{\eps_\sT^{ij} \,R_{\sT i}\,k_{\sT j}}{M_1\,M_2}\; 
\frac{1}{2z_h} \int [d\sig_h d\tau_h ] \  \left[-B_8 \right]  
\label{eq:projg-g5} \\[2mm]
\Delta^{[i\sig^{i-} \g_5]}(z_h,\xi,{\vec k}_\sT^{\,2},
{\vec R}_\sT^{\,2},{\vec k}_\sT \cdot {\vec R}_\sT) &\equiv& 
{\epsilon_\sT^{ij}R_{\sT j}\over M_1+M_2}\, 
H_1^{\newangle}(z_h,\xi,{\vec k}_\sT^{\,2},{\vec R}_\sT^{\,2},
   {\vec k}_\sT \cdot {\vec R}_\sT)
+{\epsilon_\sT^{ij}k_{\sT j}\over M_1+M_2}\,
H_1^\perp(z_h,\xi,{\vec k}_\sT^{\,2},{\vec R}_\sT^{\,2},
   {\vec k}_\sT \cdot {\vec R}_\sT) \nn\\
&=& {\epsilon_\sT^{ij}R_{\sT j}\over M_1+M_2}\, 
\frac{1}{2z_h}\int [d\sig_h d\tau_h] \, 
\left[-B_5\left(\frac{M_1+M_2}{z_hM_1}\right)
      +B_6\left(\frac{M_1+M_2}{z_hM_2}\right)-B_7\right]\nn\\
&& \hspace{-16mm}
{}+{\epsilon_\sT^{ij}k_{\sT j}\over M_1+M_2}\,
\frac{1}{2z_h}\int [d\sig_h d\tau_h] \, 
\left[B_5\,\xi\left(\frac{M_1+M_2}{M_1}\right)+B_6\,(1-\xi) 
\left(\frac{M_1+M_2}{M_2}\right)\right] \; , \label{eq:projsigi-g5} 
\end{eqnarray}  
where $\epsilon_\sT^{\mu\nu}\equiv\epsilon^{-+\mu\nu}$ (such that $i,j$ are
transverse indices) and  
\begin{equation}
\int [d\sig_h d\tau_h] \equiv \int d\sig_h \, d\tau_h \;
\delta\left(\tau_h+{\vec k}_\sT^{\,2}-\frac{\sig_h}{z_h}+\frac{M_h^2}{z_h^2}
\right)  \;.
\end{equation}
Transverse 4-vectors are defined as 
$a_\sT^\mu=g_\sT^{\mu\nu}\,a_\nu=[0,0,\vec a_\sT]$ (with 
$g_\sT^{\mu\nu}=g^{\mu\nu}-n_+^\mu n_-^\nu-n_+^\nu n_-^\mu$).

\begin{figure}[h]
\begin{center}
\psfig{file=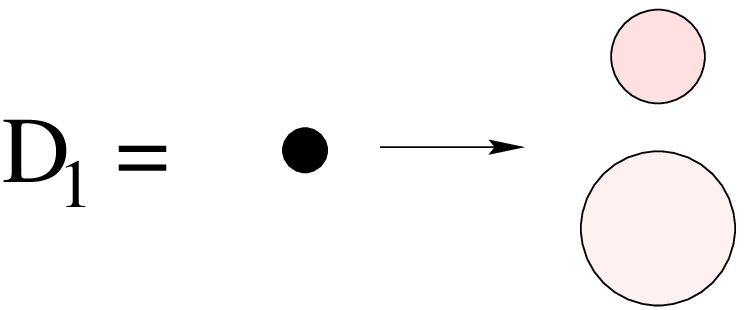, width=5cm} \\
\vspace{1cm}
\psfig{file=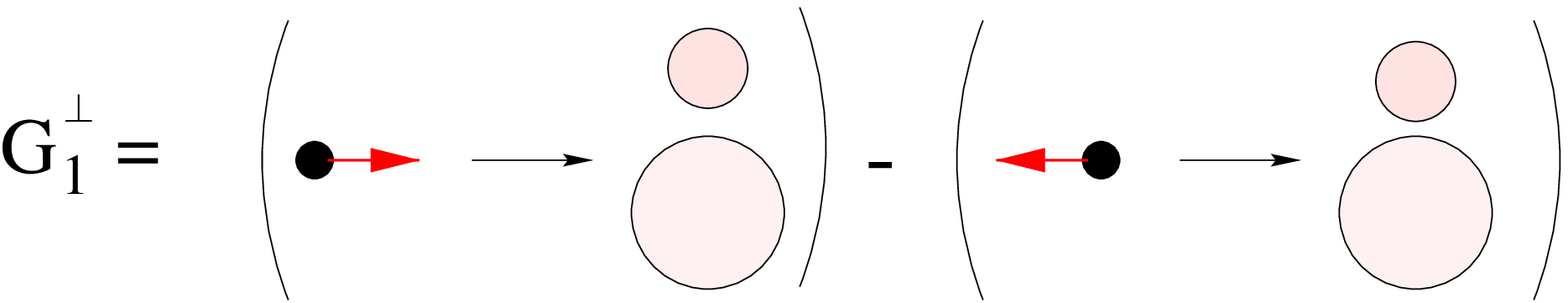, width=12cm} \\
\vspace{1cm}
\psfig{file=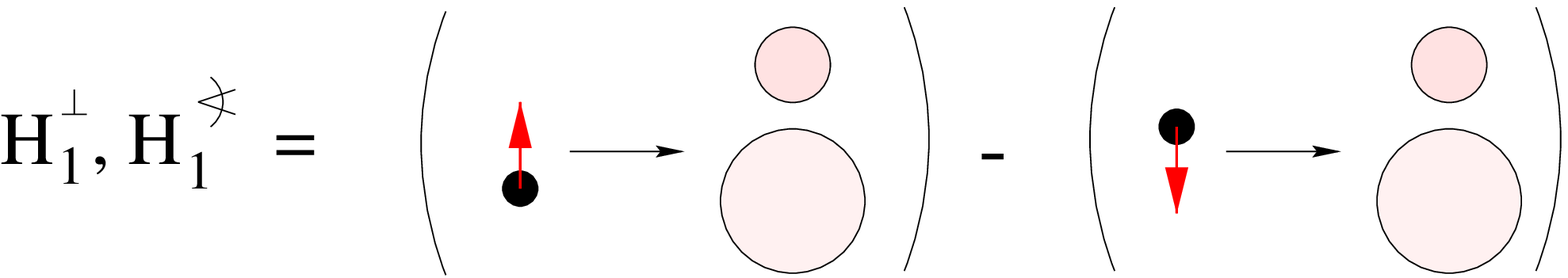, width=12cm} 
\end{center}
\caption{Probabilistic interpretation for the leading order FF arising in 
the decay of a current quark into a pair of unpolarized hadrons.}
\label{fig5}
\end{figure}

The functions $D_1$, $G_1^\perp$, $H_1^{\newangle}$, $H_1^\perp$ are the 
FF that arise to leading order in $1/Q$ for the 
fragmentation of a current quark into two unpolarized hadrons inside the 
same jet. The different Dirac structures used in the projections are 
related to different spin states of the fragmenting quark and lead to a 
nice probabilistic interpretation~\cite{piet96}. As illustrated in 
Fig.~\ref{fig5}, $D_1$ is the probability for an unpolarized quark to 
produce a pair of unpolarized hadrons; $G_1^\perp$ is the difference of 
probabilities for a longitudinally polarized quark with opposite 
chiralities to produce a pair of unpolarized hadrons; $H_1^{\newangle}$ 
and $H_1^\perp$ both are differences of probabilities for a transversely 
polarized quark with opposite spins to produce a pair of unpolarized 
hadrons. 

The interference functions $G_1^\perp$, $H_1^{\newangle}$ and 
$H_1^\perp$ are (naive) 
``T-odd''. In fact, the probability for an anyway polarized quark with 
observed transverse momentum to fragment into unpolarized hadrons is 
nonvanishing only if there are residual interactions in the final state. 
In this case, constraints from time-reversal invariance cannot be applied, 
i.e.\ the condition (\ref{eq:todd}) does not apply, and indeed the 
projections (\ref{eq:projg-g5}),(\ref{eq:projsigi-g5}) are nonvanishing. 
A measure of these functions would directly give the size and importance 
of such FSI inside the jet. 

$G_1^\perp$ is chiral even; it has a counterpart in the FF for one-hadron
semi-inclusive production. In that case, from the $\Delta^{[\g^-]}$ 
projection a ``T-odd'' FF arises, named $D_{1T}^{\perp}$, which describes the 
probability for an unpolarized quark with observed transverse momentum to 
fragment in a transversely polarized hadron~\cite{piet96}. It is known 
also in a different context~\cite{bible} that the similarity is recovered 
by substituting an axial vector (the hadron transverse spin) with a vector 
(the momentum of a second detected hadron) and by balancing this change in 
parity with the introduction of the quark polarization. 

The functions $H_1^{\newangle}$ and $H_1^\perp$ are chiral odd and can, 
therefore, be identified as the chiral partners needed to access the 
transversity $h_1$, as it will be shown in Sec.~\ref{sec:Xsections}. Given 
their probabilistic interpretation, they can be 
considered as a sort of ``double'' Collins effect~\cite{coll}. They differ 
just by geometrical weighting factors that are selectively sensitive 
either to the relative momentum of the final hadrons $({\vec R}_\sT)$ or 
to the relative orientation of the total pair momentum with respect to the 
jet axis (${\vec k}_\sT$, see also Fig.~\ref{fig4}).

\section{Azimuthal asymmetries in two-hadron inclusive DIS}
\label{sec:Xsections}

We discuss the two-hadron inclusive DIS cross section for the general
situation of any two unpolarized hadrons produced in the quark current jet 
as an example for a hard process in which interference FF can be 
measured. We demonstrate briefly that asymmetry measurements allow for 
the isolation of each individual interference FF convoluted with a specific 
DF. Only leading order $(1/Q)^0$ effects are discussed and we do not consider
QCD corrections, i.e.\ we focus on tree-level $(\alpha_s)^0$.

\begin{figure}[hbtp]
\begin{center} 
\vspace*{-5mm}
\psfig{file=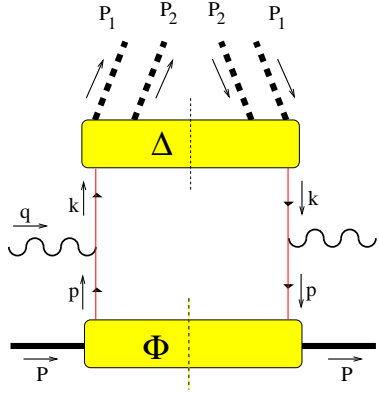, width=5cm}
\end{center} 
\caption{\label{leadDiag} Quark diagram contributing in leading order to 
two-hadron inclusive DIS when both hadrons are in the same quark current 
jet. There is a similar diagram for anti-quarks.}
\end{figure}

To leading order the hadron tensor of the process, including quarks and
anti-quarks, is 
(see Fig.~\ref{leadDiag} for the definition of momenta)
\begin{eqnarray} 
2M\, {\cal W}^{\mu\nu}&=& 
\int dp^-\,dk^+\,d^{\,2}{\vec p}_\sT^{}\,d^{\,2}{\vec k}_\sT^{}\; 
\delta^2\!\left({\vec p}_\sT^{}+{\vec q}_\sT^{}-{\vec k}_\sT^{}\right) \nn\\
&& \hspace{26mm}\times
 \mbox{Tr}\big[ 
\; \Phi(p;P,S) \; \gamma^\mu \; \Delta(k;P_1,P_2) \; \gamma^\nu \; \big]
\Big|_{\tiny\begin{array}{l}p^+ = x P^+ \\ k^- = P_h^-/z_h\end{array}}
+ \left(\begin{array}{c} 
q\leftrightarrow -q \\ \mu \leftrightarrow \nu
\end{array} \right) \; ,
\end{eqnarray} 
where $M$ is the target hadron mass.

The quark-quark correlation functions $\Phi$ for the (spin-1/2) target 
hadron is defined (in the light-cone gauge) as
\begin{equation} 
\Phi_{mn}(p;P,S) = \int \frac{d^{\,4\!}\zeta}{(2\pi)^4}\;
e^{ip \cdot \zeta} \;
\langle{P,S\vert\,\overline{\psi}_n (0) \psi_m (\zeta) \,\vert P,S}\rangle\;.
\end{equation} 
Using Lorentz invariance, hermiticity, and parity invariance, the 
(partly integrated) correlation function $\Phi$ is parametrized in terms
of DF as
\begin{eqnarray}
\Phi(x,{\vec p}_\sT) &\equiv & 
\left. \int dp^-\ \Phi(p;P,S) \right|_{p^+ = x P^+}
\!\!\!\!=\frac{M}{2P^+}\,\Biggl\{
f_1\, \frac{\Pslash}{M} + 
f_{1T}^\perp\, \epsilon_{\mu \nu \rho \sigma}\gamma^\mu 
\frac{P^\nu p_\sT^\rho S_{\sT}^\sigma}{M^2}
- \left(\lambda\,g_{1L}
        +\frac{({\vec p}_\sT\cdot{\vec S}_{\sT})}{M}\,g_{1T}\right)
  \frac{\Pslash \gamma_5}{M}
\nn \\[2 mm] 
&&
- h_{1T}\,\frac{i\sigma_{\mu\nu}\gamma_5 S_{\sT}^\mu P^\nu}{M}
- \left(\lambda\,h_{1L}^\perp
        +\frac{({\vec p}_\sT\cdot{\vec S}_{\sT})}{M}\,h_{1T}^\perp\right)\,
  \frac{i\sigma_{\mu\nu}\gamma_5 p_\sT^\mu P^\nu}{M^2}
+ h_1^\perp \, \frac{\sigma_{\mu\nu} p_\sT^\mu
P^\nu}{M^2}\Biggl\},
\end{eqnarray}
where the DF depend on the usual invariant $x=Q^2/(2P\cdot q)$ and the quark 
transverse momentum ${\vec p}_\sT$~\cite{piet96,danielpiet}. The 
polarization state of the target is fully specified by the light-cone 
helicity $\lambda = M\,S^+/P^+$ and the transverse spin $\vec S_\sT$ of the 
target hadron. The quark-quark correlation function $\Delta$ has the
structure discussed in Sec.~\ref{sec:four}. 

The definitions of DF and FF are given in a
reference frame where the target hadron momentum $P$ and the momentum of the 
produced hadron pair $P_h$ have no transverse 
components, i.e.~$\vec P_\sT=\vec P_{h\sT}=0$ and $\vec q_\sT\ne 0$ 
(see Appendix~\ref{app:kin}).

\begin{figure}[htb]
\begin{center}
\psfig{file=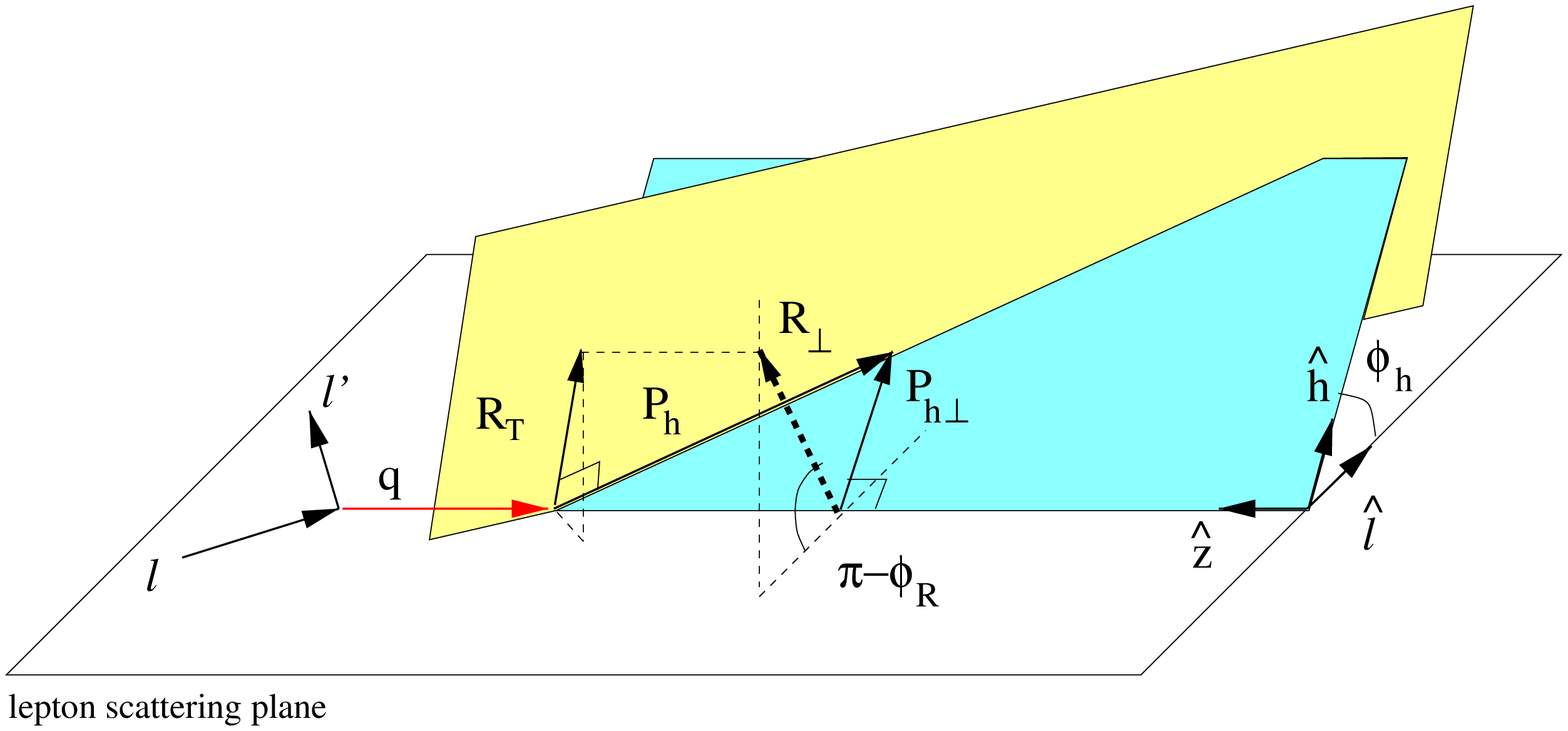, width=13cm}
\end{center}
\caption{\label{fig7} Kinematics for two-hadron inclusive 
leptoproduction.
The lepton scattering plane is determined by the momenta $l$, $l^\prime$. The 
momentum $P_h=P_1+P_2$ of the hadron pair represents the intersection between
the lightly shaded hadron-pair plane (containing $\vec R_\sT$) and the shaded
plane that defines the azimuthal dependence of the pair 
emission. $\vec R_{\perp}$ and $\vec P_{h\, \perp}$ lie in a plane 
perpendicular to the
scattering one, where $P,q$ are collinear 
($\vec P_{\perp} = \vec q_{\perp} = 0$).}
\end{figure}

For the analysis of the cross section of the full DIS process, however, a 
different frame turns out to be more useful. Angular dependences are 
conveniently expressed in a frame where the target hadron momentum $P$ and 
the photon momentum $q$ are collinear: transverse components in this frame 
are indicated with a $\perp$ subscript, thus $\vec P_\perp=\vec q_\perp=0$ 
and $\vec P_{h\perp}\ne 0$ (see Fig.~\ref{fig7} and 
Refs.~\cite{piet96,danielpiet} for more details about the kinematics). The 
cross sections should be kept differential in $d^{\,2\!} \vec P_{h\perp}$ 
for which the relation $\vec q_\sT=-\vec P_{h\perp}/z_h$ holds.
 
All azimuthal angles are defined with respect 
to $\hat l_\perp^\mu$, which is the normalized perpendicular part of 
the incoming lepton momentum $l$ such that, for a generic 
perpendicular 4-vector $a$, 
\begin{eqnarray}
\hat{l}_\perp \cdot a_\perp &=& - |{\vec a}_\perp | \cos \phi_a, \\
\epsilon^{\mu\nu}_\perp \hat{l}_{\perp\mu} a_{\perp\nu}&=& |{\vec a}_\perp | 
\sin \phi_a,
\end{eqnarray}
where $\epsilon^{\mu\nu}_\perp=\epsilon^{\mu\nu\rho\sigma}\,q_\rho\,P_\sigma
/(P\cdot q)$. Frequently, we will use the normalized perpendicular vector 
$\hat h^\mu=P_{h\perp}^\mu/|{\vec P}_{h\perp}|
=g_\perp^{\mu\nu}P_{h\nu}/|{\vec P}_{h\perp}|$ (with
$g^{\mu\nu}_\perp=g^{\mu\nu}-(Q^2P^\mu P^\nu)/(P\cdot q)$).

A comment has to be made about the definition of the perpendicular vectors
$R_\perp$ and $S_\perp$, which are obtained from the transverse 
4-vectors $R_\sT$ and $S_\sT$ by transforming with an appropriate Lorentz 
boost to the frame where $P$ and $q$ are collinear. Only the components 
which are perpendicular in the new frame are kept. Technically this 
procedure amounts to 
$R_\perp^\mu=g_\perp^{\mu\nu}\; R_{\sT\, \nu}
=g_\perp^{\mu\nu}\;\left(g_{\nu\rho}^\sT \; R^\rho \right)$ and similar 
for $S_\perp$.

The differential cross section for the process under consideration is obtained
by contraction of the hadron tensor with the standard lepton tensor, and
involves convolution integrals of DF and FF of the generic 
form~\footnote{Note that the transverse 
components ${\vec p}_\sT$ and ${\vec k}_\sT$ defined in the frame of 
Appendix~\ref{app:kin} are integration variables in the convolution, and 
thus need not to be reexpressed in the perpendicular frame.}
\begin{equation} 
{\cal F}\left[w({\vec p}_\sT^{},{\vec k}_\sT^{})\; f\, D\right]\equiv \;
\sum_a e_a^2\;
\int d^{\,2}{\vec p}_\sT^{}\; d^{\,2}{\vec k}_\sT^{}\;
\delta^2 ({\vec p}_\sT^{}+{\vec q}_\sT^{}-{\vec k}_\sT^{}) \;
w({\vec p}_\sT^{},{\vec k}_\sT^{})\;
f^a(x,{\vec p}_\sT)\,D^a(z_h,\xi,{\vec k}_\sT^{\,2},{\vec R}_\sT^{\,2},
   {\vec k}_\sT \cdot {\vec R}_\sT) \;,
\end{equation}
where $w({\vec p}_\sT^{},{\vec k}_\sT^{})$ is a weight function and the sum 
runs over all quark (and anti-quark) flavors, with $e_a$ the electric charges 
of the quarks. In the cross sections we 
will encounter the following functions of the lepton invariant 
$y=(P \cdot q)/ (P \cdot l) \approx q^-/l^-$:
\begin{equation} 
A(y) = \left(1-y+\frac{1}{2}y^2\right), \quad
B(y) = (1-y), \quad
C(y) = y (2-y).
\end{equation} 

In the following we discuss some 
particularly interesting terms of the cross section for special cases
of beam and target polarizations. The full cross section is listed 
in Appendix~\ref{App:xsect}. However, parts that cancel in taking 
differences of cross
sections with reversed polarizations are not shown. 

With a polarized beam ($L$), with helicity $\lambda_e$, scattering on an 
unpolarized target ($O$), the cross section 
\begin{equation} 
\label{xsectLO}
\frac{d\sigma(\ell H\to \ell' h_1h_2 X)_{LO}}
     {d\Omega\,dx\,dz_h\,d\xi\,d^{\,2\!}{\vec P}_{h\perp}^{}\,
                               d^{\,2\!}{\vec R}_\perp^{}}
\propto\Bigg\{ \ldots 
-\lambda_e\;|{\vec R}_\perp|\;\;C(y)\;\sin(\phi_h-\phi_R)\;
   {\cal F}\left[\,{\vec{\hat h}}\!\cdot \!\vec k_\sT^{}\,
     \frac{f_1 \, G_1^{\perp}}{2M_1M_2}\right]
\Bigg\}
\end{equation} 
is sensitive to a convolution of the unpolarized DF $f_1$ with 
$G_1^{\perp}$. The azimuthal angles $\phi_h, \phi_R$ are shown in
Fig.~\ref{fig7}, while $\Omega$ represents the solid angle of the scattered
lepton. Repeating the measurement with reversed beam helicity $\lambda_e$ 
and taking the difference of the cross sections singles out the term of
interest. 

A very similar term occurs in the cross section for an unpolarized beam 
($O$) scattering on a longitudinally polarized target ($L$) 
\begin{eqnarray}
\label{xsectOL}
\frac{d\sigma(\ell H\to \ell' h_1 h_2 X)_{OL}}
     {d\Omega\,dx\,dz_h\,d\xi\,d^{\,2\!}{\vec P}_{h\perp}^{}\,
                               d^{\,2\!}{\vec R}_\perp^{}}
&\propto&\Bigg\{ \ldots
{}-\lambda\;|{\vec R}_\perp|\;A(y)\;\sin(\phi_h-\phi_R)\;
   {\cal F}\left[\,{\vec{\hat h}}\!\cdot \!\vec k_\sT^{}\,
     \frac{g_{1} \, G_1^{\perp}}{M_1M_2}\right] +\ldots
\Bigg\} \; ,
\end{eqnarray} 
where the FF $G_1^{\perp}$ is convoluted with the polarized DF $g_1$. The
azimuthal angular dependence is the same as before. For this experiment, 
reversing the polarization of the 
target is not sufficient to single out the interesting term, since the full 
cross section (cf.~Appendix~\ref{App:xsect}) contains more contributions 
which do not cancel in the difference. One has to analyze the azimuthal 
angular dependence, which is unique for each contributing term. 

With a transversely polarized target ($T$) and unpolarized beam ($O$), the
cross section contains the contributions
\begin{eqnarray}
\label{eq:xsectOT}
\frac{d\sigma(\ell H\to \ell' h_1 h_2 X)_{OT}}
     {d\Omega\,dx\,dz_h\,d\xi\,d^{\,2\!}{\vec P}_{h\perp}^{}\,
                               d^{\,2\!}{\vec R}_\perp^{}}
&\propto&\Bigg\{ \ldots
{}+|{\vec S}_\perp|\;B(y)\;\sin(\phi_h+\phi_{S})
   {\cal F}\left[\,{\vec{\hat h}}\!\cdot \!\vec k_\sT^{}\,
     \frac{h_1 \, H_1^{\perp}}{M_1+M_2}\right]\nn\\
&&
{}+|{\vec S}_\perp|\;|{\vec R}_\perp|\;B(y)\;\sin(\phi_R+\phi_{S})\;
   {\cal F}\left[\frac{h_1 \, H_1^{\newangle}}{M_1+M_2}\right]
+\ldots\Bigg\} \; ,
\end{eqnarray} 
which involve the transversity DF $h_1$, and the FF $H_1^{\perp}$ and 
$H_1^{\newangle}$, respectively. The experimental situation is analogous to
the one proposed to access the transversity $h_1$ in one-hadron inclusive DIS
via the so-called Collins effect~\cite{coll}. In the process under
consideration, two contributions of similar kind arise which can be analyzed
separately using their different kinematical signatures. In fact, asymmetry
measurements can firstly be done, that isolate both 
contributions in Eq.~(\ref{eq:xsectOT}). Then, the analysis of the asymmetry 
produced by interchanging the relative position of the hadron pair (i.e., by 
flipping $\vec R$ by $180^{\rm o}$) isolates the convolution containing 
$H_1^{\newangle}$. This second term is interesting, since the absence of a 
$\vec k_\sT-$ (or $\vec p_\sT-$) dependent weight factor in the convolution 
integral allows for a nonvanishing integration 
of the differential cross section over $d^{\,2\!} \vec P_{h\perp}$, such 
that the the convolution turns into a product of DF and FF
\begin{eqnarray} 
\int d^{\,2\!} \vec P_{h\perp} \;
{\cal F}\left[\frac{h_1 \, H_1^{\newangle}}{M_1+M_2}\right]&=&
\frac{z_h^2}{M_1+M_2} \; \sum_a e_a^2 \;
\int d^{\,2}{\vec p}_\sT^{} \; h_1^a(x,{\vec p}_\sT)\;\;
\int d^{\,2}{\vec k}_\sT^{}\;
H_1^{\newangle\,a}(z_h,\xi,{\vec k}_\sT^{\,2},{\vec R}_\sT^{\,2},
   {\vec k}_\sT \cdot {\vec R}_\sT)\nn\\
&=&
\frac{z_h^2}{M_1+M_2} \; \sum_a e_a^2 \;\;\; h_1^a(x)  \;\;\;
H_1^{\newangle\,a}(z_h,\xi,{\vec R}_\sT^{\,2}).
\end{eqnarray} 
The corresponding experimental situation is favorable, since less
kinematical variables have to be determined and the quantity of interest
depends on $z_h$, $\xi$ and ${\vec R}_\sT^{\,2}$ only. Note, however, that 
terms in 
$H_1^{\newangle}(z_h,\xi,{\vec k}_\sT^{\,2},{\vec R}_\sT^{\,2},
{\vec k}_\sT \cdot {\vec R}_\sT)$ with odd powers 
of ${\vec k}_\sT$, like for 
instance ${\vec k}_\sT\cdot{\vec R}_\sT$, do not survive the symmetric 
integration over $d^{\,2}{\vec k}_\sT^{}$.

\section{Summary and Outlooks}
\label{sec:summary}

In this paper we have investigated the general properties of 
interference fragmentation functions (FF) that arise from the distribution 
of two hadrons produced in the same jet in the current fragmentation 
region of a hard process, e.g., in two-hadron inclusive lepton-nucleon 
scattering.

Naive ``T-odd'' FF generally arise because the existence of 
Final State Interactions (FSI) prevents constraints from time-reversal 
invariance to be applied. This class of FF is interesting for several
reasons: obviously, being FF the ``decay-channel'' partners of the 
distribution functions (DF), they can give information on the 
parton structure of 
hadrons that are not available as targets; they are directly related to FSI 
and, therefore, give access to exploration of mechanisms for residual 
interactions inside jets; 
finally, a subset of these FF is chiral odd and represents the needed 
partner to isolate the quark transversity distribution, which is required 
to complete the picture of the quark structure of hadrons at leading 
order but, at the same time, is presently completely unknown due to its 
chiral-odd nature.

The presence of FSI allows that in the fragmentation process there are at 
least two competing channels interfering through a nonvanishing phase. 
However, it has been shown that this is not enough to generate ``T-odd'' 
FF. A genuine difference in the Lorentz structure of the vertices 
describing the fragmentation is needed. This argument naturally selects 
the considered process, namely two-hadron emission inside the same jet in
semi-inclusive DIS, as the simplest scenario for modelling the
fragmentation. 

To leading order, four FF arise, which have a nice probabilistic
interpretation. They can be grouped in three classes according to the
polarization of the fragmenting quark. We have studied, in particular, 
those FF for quarks polarized longitudinally $(G_1^{\perp})$ and 
transversely $(H_1^{\perp}$, $H_1^{\newangle})$, that evolve fragmenting 
into a pair of unpolarized hadrons. These FF are naive ``T-odd''. The 
former is chiral even, while the latter are chiral odd and represent a 
sort of ``double'' Collins effect. 

Asymmetry measurements in two-hadron inclusive DIS that allow isolation of 
each individual FF are possible and are described in 
Sec.~\ref{sec:Xsections}. Both $H_1^{\newangle}$, $H_1^{\perp}$ enter the 
cross section in convolutions with the transversity distribution $h_1$ to 
leading order, thus permitting its extraction from a measurement of 
deep-inelastic scattering on a transversely polarized nucleon target
induced by an unpolarized beam. In particular, the term of the cross
section involving $H_1^{\newangle}$ survives the integration over the 
transverse momentum of the fragmenting quark and results in a deconvolution of 
the FF and the transversity distribution $h_1$.

\acknowledgements{This work is part of the TMR program ERB FMRX-CT96-0008.
Interesting and fruitful discussion with D.~Boer and P.~Mulders are 
greatly acknowledged.}

\appendix
\section{}
\label{app:kin}

We use two dimensionless light-like vectors $n_+$ and $n_-$ (satisfying
$n_+^2=n_-^2=0$ and $n_+\cdot n_-=1$) to decompose a 4-vector $a$ in
its light-cone components 
$a^\pm=\left(a^0\pm a^3\right)/\sqrt{2}=a_\cdot n_\mp$ and a 
two-dimensional transverse vector $\vec a_\sT$. To display an explicit
parametrization of 4-vectors we use the notation
$a^\mu=\left[a^-,a^+, {\vec a}_\sT\right]$.

Generally, the definitions of distribution and fragmentation functions are
given in a reference frame, where the hadron momentum has no transverse
momentum. For the case of two-hadron production in the same jet, the 
corresponding
frame is the one where the sum $P_h=P_1+P_2$ has zero transverse momentum
\begin{equation}
P_h=P_1+P_2=\lcvec{P_h^-}{P_h^+}{{\vec 0}_\sT} \;.
\end{equation} 
When we discuss the two-hadron fragmentation as a part of the full
two-hadron inclusive DIS process, as done in Sec.~\ref{sec:Xsections},
definitions are given in the frame where $P_h$ and the target hadron momentum
$P$ are collinear, i.e.~$\vec P_{h\sT}=\vec P_\sT=0$.  

The quark-quark correlation $\Delta$ depends on the three 4-momenta 
$k,P_1,P_2$, from which the following light-cone 
fractions can be defined
\begin{equation}
z_h=\frac{{P_h}^-}{k^-}=\frac{P_1^-+P_2^-}{k^-}
=\frac{P_1^-}{k^-}+\frac{P_2^-}{k^-} =z_1+z_2 \; .
\label{eq:z}
\end{equation} 
An explicit parametrization for the three momenta external to $\Delta$ is
\begin{eqnarray}
k&=&
\lcvec{\frac{P_h^-}{z_h}}{z_h\frac{k^2+{\vec k}_\sT^{\,2}}{2P_h^-}}
      {{\vec k}_\sT}
\nn \\
P_1&=&
\lcvec{P_h^-\frac{z_1}{z_h}}{\frac{z_h(M_1^2+{\vec R}_\sT^{\,2})}
      {2\,z_1 P_h^-}}
{{\vec R}_\sT} \nn \\
P_2&=&
\lcvec{P_h^-\frac{z_2}{z_h}}{\frac{z_h(M_2^2+{\vec R}_\sT^{\,2})}
      {2\,z_2 P_h^-}}
{-{\vec R}_\sT} \; ,
\label{eq:vectors}
\end{eqnarray}
with $R\equiv(P_1-P_2)/2$ being (half of) the relative momentum 
between the hadron pair. Then, the invariants defined in 
Eq.~(\ref{eq:invariants}) become
\begin{mathletters}
\label{explinv}
\begin{eqnarray}
\tau_h &=& k^2 \nn \\
\sig_h &=& 2k\cdot (P_1+P_2)=
2k\cdot P_h =\left\{\frac{M_1^2+{\vec R}_\sT^{\,2}}{z_1}+
\frac{M_2^2+{\vec R}_\sT^{\,2}}{z_2}\right\}+z_h\,(\tau_h+{\vec k}_\sT^{\,2})
\label{eq:sigh}\\
\sig_d &=& 2k\cdot (P_1-P_2)=
4k\cdot R =\left\{\frac{M_1^2+{\vec R}_\sT^{\,2}}{z_1}
-\frac{M_2^2+{\vec R}_\sT^{\,2}}{z_2}\right\}
+(z_1-z_2)(\tau_h+{\vec k}_\sT^{\,2})
-4\,{\vec k}_\sT\cdot{\vec R}_\sT \label{eq:sigd} \\
M_h^2 &=& P_h^2 = 2\,P_h^+\,P_h^-=
z_h\left\{\frac{M_1^2+{\vec R}_\sT^{\,2}}{z_1}
+\frac{M_2^2+{\vec R}_\sT^{\,2}}{z_2}\right\} \; . 
\label{eq:mh}
\end{eqnarray} 
\end{mathletters}

Alternatively, by defining $\xi=z_1/z_h$ the external 4-momenta are
\begin{eqnarray}
k&=&
\lcvec{\frac{P_h^-}{z_h}}{z_h\frac{k^2+{\vec k}_\sT^{\,2}}{2P_h^-}}
      {{\vec k}_\sT} 
\nn\\
P_1&=&
\lcvec{\xi\,P_h^-}{\frac{M_1^2+{\vec R}_\sT^{\,2}}{2\,\xi\,P_h^-}}
      {{\vec R}_\sT}
\nn\\
P_2&=&
\lcvec{(1-\xi)\,P_h^-}{\frac{M_2^2+{\vec R}_\sT^{\,2}}{2\,(1-\xi)\,P_h^-}}
{-{\vec R}_\sT} 
\label{eq:vectors2}
\end{eqnarray}
and the invariants read
\begin{mathletters}
\label{explinv2}
\begin{eqnarray}
\tau_h &=& k^2 \nn \\
\sig_h &=& 2k\cdot P_h =\left\{\frac{M_1^2+{\vec R}_\sT^{\,2}}{z_h\,\xi}
+\frac{M_2^2+{\vec R}_\sT^{\,2}}{z_h\,(1-\xi)}\right\}
+z_h\,(\tau_h+{\vec k}_\sT^{\,2})
\label{eq:sigh2} \\
\sig_d &=& 2k\cdot (P_1-P_2) =\left\{\frac{M_1^2+{\vec R}_\sT^{\,2}}{z_h\,\xi}
-\frac{M_2^2+{\vec R}_\sT^{\,2}}{z_h\,(1-\xi)}\right\}+z_h(2\xi-1)(\tau_h+
{\vec k}_\sT^{\,2})-4\,{\vec k}_\sT\cdot{\vec R}_\sT \label{eq:sigd2} \\
M_h^2 &=& P_h^2 = 2\,P_h^+\,P_h^-=\left\{\frac{M_1^2+{\vec R}_\sT^{\,2}}{\xi}
+\frac{M_2^2+{\vec R}_\sT^{\,2}}{1-\xi}\right\} \; . 
\label{eq:mh2}
\end{eqnarray}
\end{mathletters}
Eq.~(\ref{eq:mh2}) can also be expressed as
\begin{equation}
{\vec R}_\sT^{\,2}=\xi\,(1-\xi)\,M_h^2-(1-\xi)\,M_1^2-\xi\,M_2^2 \; .
\label{eq:pt2}
\end{equation}

\section{}
\label{App:xsect}

In this Appendix we list the full leading order cross section for two-hadron
inclusive DIS. It is shown splitted in parts for unpolarized ($O$) or
longitudinally polarized ($L$) lepton beam, and unpolarized ($O$),
longitudinally ($L$) or transversely ($T$) polarized hadronic target.
A kinematic overall factor, which cancels in any asymmetries, is omitted.
Parts that cancel in taking differences of cross sections with reversed
polarizations, are not shown.
\begin{eqnarray}
\label{app:xsectOO}
\frac{d\sigma(\ell H\to \ell' h_1h_2 X)_{OO}}
     {d\Omega\,dx\,dz_h\,d\xi\,d^{\,2\!}{\vec P}_{h\perp}^{}
\,d^{\,2\!}{\vec R}_\perp^{}}
&\propto&
\Bigg\{ 
{}A(y)\;{\cal F}\left[f_1 \, D_1\right]
{}-|{\vec R}_\perp|\;B(y)\;\cos(\phi_h+\phi_R)\;
   {\cal F}\left[\,{\vec{\hat h}}\!\cdot \!\vec p_\sT^{}\,
     \frac{h_1^{\perp} \, H_1^{\newangle}}{M(M_1+M_2)}\right]\nn\\
&&
{}-B(y)\;\cos(2\phi_h)\;
   {\cal F}\left[\left(2\,{\vec{\hat h}}\!\cdot\!\vec p_\sT^{}\,
                        \,{\vec{\hat h}}\!\cdot \! \vec k_\sT^{}\,
                       -\,\vec p_\sT^{}\!\cdot \! \vec k_\sT^{}\,\right)
     \frac{h_1^{\perp} \, H_1^{\perp}}{M(M_1+M_2)}\right]
\Bigg\}
\end{eqnarray} 

\begin{equation} 
\label{app:xsectLO}
\frac{d\sigma(\ell H\to \ell' h_1h_2 X)_{LO}}
     {d\Omega\,dx\,dz_h\,d\xi\,d^{\,2\!}{\vec P}_{h\perp}^{}\,
d^{\,2\!}{\vec R}_\perp^{}}
\propto\Bigg\{
\ldots
{}-\lambda_e\;|{\vec R}_\perp|\;\;C(y)\;\sin(\phi_h-\phi_R)\;
   {\cal F}\left[\,{\vec{\hat h}}\!\cdot \!\vec k_\sT^{}\,
     \frac{f_1 \, G_1^{\perp}}{2M_1M_2}\right]
\Bigg\}
\end{equation} 

\begin{eqnarray}
\label{app:xsectOL}
\frac{d\sigma(\ell H\to \ell' h_1 h_2 X)_{OL}}
     {d\Omega\,dx\,dz_h\,d\xi\,d^{\,2\!}{\vec P}_{h\perp}^{}\,
d^{\,2\!}{\vec R}_\perp^{}}
&\propto&\Bigg\{ 
\ldots
{}-\lambda\;|{\vec R}_\perp|\;A(y)\;\sin(\phi_h-\phi_R)\;
   {\cal F}\left[\,{\vec{\hat h}}\!\cdot \!\vec k_\sT^{}\,
     \frac{g_{1} \, G_1^{\perp}}{M_1M_2}\right] \nn\\
&&
{}+\lambda\;B(y)\;\sin(2\phi_h)\;
   {\cal F}\left[\left(2\,{\vec{\hat h}}\!\cdot \!\vec p_\sT^{}\,
                        \,{\vec{\hat h}}\!\cdot \!\vec k_\sT^{}\,
                       -\,\vec p_\sT^{}\!\cdot \!\vec k_\sT^{}\,\right)
     \frac{h_{1L}^{\perp} \, H_1^{\perp}}{M(M_1+M_2)}\right]\nn\\
&&
{}+\lambda\;|{\vec R}_\perp|\;B(y)\;\sin(\phi_h+\phi_R)\;
   {\cal F}\left[\,{\vec{\hat h}}\!\cdot \!\vec p_\sT^{}\,
     \frac{h_{1L}^{\perp} \, H_1^{\newangle}}{M(M_1+M_2)}\right]
\Bigg\}
\end{eqnarray} 

\begin{eqnarray}
\label{app:xsectOT}
\frac{d\sigma(\ell H\to \ell' h_1 h_2 X)_{OT}}
     {d\Omega\,dx\,dz_h\,d\xi\,d^{\,2\!}{\vec P}_{h\perp}^{}\,
d^{\,2\!}{\vec R}_\perp^{}}
&\propto&\Bigg\{
\ldots
{}+|{\vec S}_\perp|\;A(y)\;\sin(\phi_h-\phi_{S})\;
   {\cal F}\left[\,{\vec{\hat h}}\!\cdot \!\vec p_\sT^{}\,
     \frac{f_{1T}^{\perp} \, D_1}{M}\right]\nn\\
&&
{}+|{\vec S}_\perp|\;B(y)\;\sin(\phi_h+\phi_{S})
   {\cal F}\left[\,{\vec{\hat h}}\!\cdot \!\vec k_\sT^{}\,
     \frac{h_1 \, H_1^{\perp}}{M_1+M_2}\right]\nn\\
&&
{}+|{\vec S}_\perp|\;|{\vec R}_\perp|\;B(y)\;\sin(\phi_R+\phi_{S})\;
   {\cal F}\left[\frac{h_1 \, H_1^{\newangle}}{M_1+M_2}\right]\nn\\
&&
{}+|{\vec S}_\perp|\;|{\vec R}_\perp|\;A(y)\;
   \sin(\phi_h-\phi_{S})\;\cos(\phi_h-\phi_R)\;
   {\cal F}\left[\,\vec p_\sT^{}\!\cdot \! \vec k_\sT^{}\,
     \frac{g_{1T} \, G_1^{\perp}}{MM_1M_2}\right]\nn\\
&&
{}-|{\vec S}_\perp|\;|{\vec R}_\perp|\;A(y)\;
   \sin(2\phi_h-\phi_R-\phi_{S})\;
   {\cal F}\left[\,{\vec{\hat h}}\!\cdot \!\vec p_\sT^{}\,
                 \,{\vec{\hat h}}\!\cdot \!\vec k_\sT^{}\,
     \frac{g_{1T} \, G_1^{\perp}}{MM_1M_2}\right]\nn\\
&&
{}+|{\vec S}_\perp|\;B(y)\;\sin(3\phi_h-\phi_{S})\nn\\
&&\quad\times
   {\cal F}\left[\left(4\,(\!\,{\vec{\hat h}}\!\cdot \!\vec p_\sT^{}\,\!)^2
                             \,{\vec{\hat h}}\!\cdot \!\vec k_\sT^{}
                  -2\,{\vec{\hat h}}\!\cdot \!\vec p_\sT^{}\,
                    \,\vec p_\sT^{}\!\cdot \!\vec k_\sT^{}\,
        -\,\vec p_\sT^{\,2}\,{\vec{\hat h}}\!\cdot \!\vec k_\sT^{}\, \right)
     \frac{h_{1T}^{\perp} \, H_1^{\perp}}{2M^2(M_1+M_2)}\right]\nn\\
&&
{}+|{\vec S}_\perp|\;|{\vec R}_\perp|\;B(y)\;
    \sin(2\phi_h+\phi_R-\phi_{S})\nn\\
&&\qquad\times
   {\cal F}\left[\left(2(\hat{\vec h}\!\cdot\!\vec p^{}_\sT)^2
                               -\vec p_\sT^{\,2}\right)
     \frac{h_{1T}^{\perp} \, H_1^{\newangle}}{2M^2(M_1+M_2)}\right]
\Bigg\}
\end{eqnarray} 

\begin{equation}
\label{app:xsectLL}
\frac{d\sigma(\ell H\to \ell' h_1 h_2 X)_{LL}}
     {d\Omega\,dx\,dz_h\,d\xi\,d^{\,2\!}{\vec P}_{h\perp}^{}
      \,d^{\,2\!}{\vec R}_\perp^{}}
\propto\Bigg\{
\ldots
{}+\lambda_e\;\lambda\;\frac{C(y)}{2}\;
   {\cal F}\left[g_{1} \, D_1\right]
\Bigg\}
\end{equation} 

\begin{eqnarray}
\label{app:xsectLT}
\frac{d\sigma(\ell H\to \ell' h_1 h_2 X)_{LT}}
     {d\Omega\,dx\,dz_h\,d\xi\,d^{\,2\!}{\vec P}_{h\perp}^{}\,
d^{\,2\!}{\vec R}_\perp^{}}
&\propto&\Bigg\{ 
\ldots
{}+\lambda_e\;|{\vec S}_\perp|\;C(y)\;\cos(\phi_h-\phi_{S})\;
   {\cal F}\left[\,{\vec{\hat h}}\!\cdot \!\vec p_\sT^{}\,
     \frac{g_{1T} \, D_1}{2M}\right]\nn\\
&&
{}-\lambda_e\;|{\vec S}_\perp|\;|{\vec R}_\perp|\;C(y)\;
    \cos(\phi_h-\phi_{S})\;\cos(\phi_h-\phi_R)\;
   {\cal F}\left[\,\vec p_\sT^{}\!\cdot \!\vec k_\sT^{}\,
     \frac{f_{1T}^{\perp} \, G_1^{\perp}}{2MM_1M_2}\right]\nn\\
&&
{}+\lambda_e\;|{\vec S}_\perp|\;|{\vec R}_\perp|\;C(y)\;
    \cos(2\phi_h-\phi_R-\phi_{S})\;
   {\cal F}\left[\,{\vec{\hat h}}\!\cdot \!\vec p_\sT^{}\,
                 \,{\vec{\hat h}}\!\cdot \!\vec k_\sT^{}\,
     \frac{f_{1T}^{\perp} \, G_1^{\perp}}{2MM_1M_2}\right]
\Bigg\}
\end{eqnarray} 
For the definition of the various ingredients entering the cross sections,
we refer the reader to Sec.~\ref{sec:Xsections}.

\end{document}